\documentclass[twocolumn,showpacs,preprintnumbers,amsmath,amssymb]{revtex4}

\usepackage{graphicx}
\usepackage{dcolumn}
\usepackage{bm}

\begin{document}
\title{Urban Traffic Dynamics: A Scale-Free Network Perspective}
\author{Mao-Bin Hu$^{1}$}\email{humaobin@ustc.edu.cn}
\author{Wen-Xu Wang$^{2}$}
\author{Rui Jiang$^{1}$}
\author{Qing-Song Wu$^{1}$}\email{qswu@ustc.edu.cn}
\author{Bing-Hong Wang$^{2}$}
\author{Yong-Hong Wu$^{3}$}
\affiliation{$^{1}$School of Engineering Science, University of
Science and Technology of China, Hefei 230026, P.R.C \\
$^{2}$Nonlinear Science Center and Department of
Modern Physics, University of Science and Technology of China,
Hefei 230026, P.R.C\\
$^{3}$Department of Mathematics and Statistics, Curtin University of Technology,
Perth, WA6845, Australia}

\date{\today}

\begin{abstract}
This letter propose a new model for characterizing traffic dynamics 
in  scale-free networks.
With a replotted road map of cities with roads mapped to vertices and 
intersections to edges, and introducing the road 
capacity $L$ and its handling ability at intersections $C$, 
the model can be applied to urban traffic system. 
Simulations give the overall capacity of the traffic system 
which is quantified by a phase transition from
free flow to congestion. Moreover, we report the fundamental
diagram of flow against density, in which hysteresis is found,
indicating that the system is bistable in a certain range of
vehicle density. In addition, the fundamental diagram is
significantly different from single-lane traffic model and 2-D BML
model with four states: free flow, saturated flow, bistable and
jammed.
\end{abstract}

\pacs{89.40.-a, 45.70.Vn, 89.75.Hc, 05.70.Fh}

\maketitle

Traffic and transportation is nowadays one of the most important
ingredients of modern society. We rely greatly on networks such 
as communication, transportation and power systems.
Ensuring free traffic flow on these networks is therefore of great 
research interest \cite{NS,Helbing,Helbing2,Kerner,Kerner2,Kerner3,
LiXB,BML,Nagatani,CKH,Angel}.
Recently, more and more empirical evidence indicates that these 
networked systems are of small-world and scale-free structural
features \cite{Rosvall,BA,BA2,Newman}. 
And so the traffic flow on scale-free networks is being
widely investigated \cite{Sole,Arenas,Tadic,Zhao,Wang}.
In the present work, we propose a new model for the traffic 
dynamics of such networks. The potential application of 
this research will be urban (road) traffic networks. 

Previously work on urban traffic normally maps roads 
to edges and intersections to vertices.
In 1992, Biham, Middleton and Levine (BML) \cite{BML} proposed an
ideal 2-dimensional model for studying urban traffic. They used a
$N \times N$ grid to represent the urban transportation network.
Initially, cars are placed at the sites. Each car is either
East-facing or North-facing. At odd time steps, each North-facing
car moves one unit North if there is a vacant site before it. At
even time steps, East-facing cars move East in the same way. 
The rules of BML model can be considered as controlling urban 
traffic by traffic lights. 
The BML model reproduced the transition from
free-flow phase to jammed phase. Since then, many studies have
been carried out on the base of the BML model. For example, Nagatani
\cite{Nagatani} investigated the effect of car accident induced
jam on the phase transition of urban traffic; Chung et al.
\cite{CKH} investigated the effect of improperly placed traffic
lights; and recently, Angel et al. \cite{Angel} discussed the
jammed phase of the BML model.

The models mensioned above always adopt a $N \times N$ grid (or close to
that) to represent the urban traffic system. 
However, real urban traffic system is obviously much more complicated.
Perhaps, the most natural way is to map each intersection to a
vertex and each segment of road to an edge between vertices.
However, obviously, this kind of simulation is computation-consuming 
and the interaction of any two neighboring segments of a straight 
road is in most cases wiped off.

\begin{figure}
\scalebox{0.65}[0.55]{\includegraphics{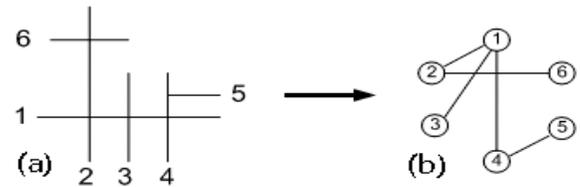}}
\caption{\label{Fig1}  Illustration of the road network mapping.
Each straight road in (a) is mapped to a vertex in (b). And
intersections are mapped to edges between the vertices.}
\end{figure}

The unique feature of our model is that we look at the urban 
traffic networks in a different point of view. 
We will create traffic flow networks 
with roads mapped to vertices and intersections to edges between vertices, 
as was inspired by the information networks suggested in \cite{Rosvall}
for an information view of a city. 
In this way(See Fig.\ref{Fig1}), the degree $k$ of vertex is the number of intersections
along the street and a major road with many minor roads connected 
to it can be seen as a major vertex. 
Empirical observations demostrate that the remapped urban networks 
exhibit scale free structural properties. 

With this new paradigm, we can look at urban traffic networks 
from a novel perspective. 
In network language, a trajectory of a vehicle can then be mapped in 
an urban traffic network map from a road (vertex) to another road (vertex) 
through a one directional channel (edge) of an intersection. 
In this work we will take this perspective to the extreme, and 
assume that the travel time/cost of just driving along a given 
road can be zero \cite{Rosvall}.
Our model is partially inspired by the work of information packet 
flow on the Internet \cite{Sole,Arenas,Tadic,Zhao,Wang}. 
The phase transition from free flow to jammed state, and the plot 
of flow against density that all have been empirically observed can be 
well reproduced by this model.

Though the previous work found that the replotted road 
networks of cities have scale-free characteristics \cite{Rosvall},
there is no well-accepted model for these networks up to now.
Without losing generality, our simulation starts from generating
the urban transportation network according to the most general 
Barab\'{a}si-Albert scale-free network model \cite{BA2}. In this
model, starting from $m_0$ fully connected vertices, one vertex
with $m$ edges is attached at each time step in such a way that
the probability $\Pi_i$ of being connected to the existing vertex
$i$ is proportional to the degree $k_i$ of the vertex, i.e.
$\Pi_i={k_i \over \Sigma_j k_j}$, where $j$ runs over all existing
vertices. The capacity of each vertex (road) is controlled by 
two parameters: (1) its maximum cars number $L$, which is proportional 
to its degree $k$ (a long road ordinarily has more intersections and 
can hold more cars): $L=\alpha \times k$; 
(2) the maximum number of cars handled per time step, which 
reflects the capability of intersections: 
$C=\beta \times L$. 
Motivated by the Internet information flow models \cite{Sole,Arenas,
Tadic,Zhao,Wang}, the system evolves in parallel according to the
following rules:

1. Add Cars - Cars are added with a given rate $R$ (cars per time
step) at randomly selected vertices and each car is given a random
destination.

2. Navigate Cars - If a car's destination is found in its
next-nearest neighborhood, its direction will be set to the
destination vertex. Otherwise, its direction will be set to a
neighboring vertex $h$ with probability:
$P_h={k^{\phi}_h \over \Sigma_i k^{\phi}_i}$.  
Here the sum runs over the neighboring vertices, and $\phi$ is an
adjustable parameter. It is assumed that the cars are unaware of
the entire network topology and only know the neighboring
vertices' degree $k_i$.

3. Cars Motion -- At each step, only at most $C$ cars can leave a
vertex (road) for other vertices and FIFO (first-in-first-out) queuing
discipline is applied at each vertex. When the queue at a selected
vertex is full, the vertex won't accept any more vehicles and the
vehicle will wait for the next opportunity. Once a car arrives at
its destination, it will be removed from the system. 

We first simulate the traffic on a network of $N=100$ vertices
(roads) with $m0=m=2$, $\alpha=5$ and $\beta=0.2$. 
This relatively small system can be seen as simulating 
the backbone of a city's urban traffic network.
The selection of $\beta$ is based on the single-road 
traffic flow theory which shows that the maximum 
flux on a highway is about $20\%$ of its maximum density
\cite{Kerner,Kerner2,Kerner3,LiXB}. For simplicity, we do not
consider the phenomenon that the flux decreases 
when the density is above $20\%$. 
The combination of $\alpha$ and $\beta$ 
can be also interpreted as: each intersection can handle one car
turning for one road at each step. 

\begin{figure}
\scalebox{0.50}[0.53]{\includegraphics{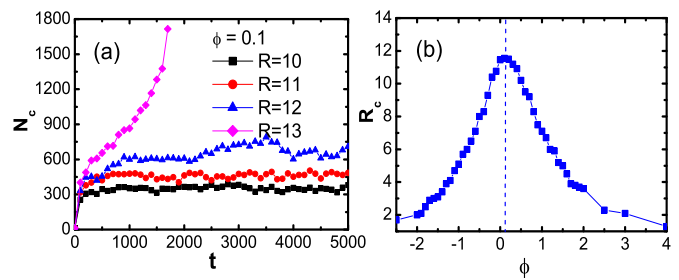}}
\caption{\label{Fig2}  (color online). The overall capacity of 
a road network with $N=100$, $m0=m=2$, $\alpha=5$ and $\beta=0.2$. 
(a) The variation of car number $N_c$ for different $R$ when 
$\phi=0.1$. $R_c(=13)$ is determined at the point where the $N_c$ 
increment rate $\omega$ increases suddenly from zero and $N_c$ 
increases rapidly towards the system's maximum car number. 
(b) The critical $R_c$ versus $\phi$. The maximum of $R_c$ 
corresponds to $\phi=0.1$ marked by a dash line. The data are 
obtained by averaging $R_c$ over 10 network realizations. }
\end{figure}

To characterize the system's overall capacity, we first
investigate the car number $N_c$ increment rate $\omega$ in the
system:
$\omega(R)=\lim_{t \rightarrow \infty} {\langle N_c(t+\Delta t)-N_c(t)
\rangle \over \Delta t}$. 
Here $\langle N_c(t+\Delta t)-N_c(t)\rangle$ takes average over time windows of
width $\Delta t$. Fig.\ref{Fig2}(a) shows the variation of
$N_c$ with different $R$ for $\phi=0.1$. One can see that there is
a critical $R_c$ ($=13$) at which $N_c$ runs quickly towards the
system's maximum car number and $\omega(R)$ increases
suddenly from zero. $\omega(R)=0$ corresponds to the cases of free
flow state, which is attributed to the balance between the number
of added and removed cars at the same time. However, if $R$
exceeds the critical value $R_c$, cars will in succession
accumulate in the system and then congestion emerges and diffuses
to everywhere. Ultimately almost no cars can arrive at their
destinations.

Evidently, $R_c$ is the onset of phase transition from free flow
to jammed state. Hence, the system's overall capacity can be
measured by the critical value of $R_c$ under which the system can
maintain its normal and efficient functioning. Fig.\ref{Fig2}(b) 
depicts the variation of $R_c$ versus $\phi$. The maximum overall
capacity occurs at $\phi=0.1$ (slightly greater than $0.0$) with
$R_c^{max}=13$.

Here we give a heuristic analysis for determining the optimal
value of $\phi$ with the maximum capacity. If we neglect
the queue length $L$ of each vertex, for $\phi=0$, cars will
move in the system nearly like random walk. There is a well-known
result from graph theory that if a particle performs a random
walk, in the limit of long times, the time the particle spends at
a given vertex is proportional to the degree $k$ of the vertex
\cite{Bollob}. One can easily find out that, the number of cars
observed averagely at a vertex is proportional to the degree
of that vertex. Meanwhile, the cars handling ability of each
vertex is assumed to be proportional to its degree. Thus in the
case of $\phi=0$, this rule produces an average effect that no
congestion occurs earlier on some vertices with particular degree
than on others. Accordingly, $\phi=0$ results in the maximum
system capacity. However, in our model, each vertex has a limited
queue length $L=\alpha\times k$ and $R$ cars are generated randomly
among all vertices at each time step, so small degree vertices are
slightly more easily congested. Therefore, for our traffic model,
a $\phi$ slightly larger than zero can enhance the system's
capacity maximally.

\begin{figure}
\scalebox{0.50}[0.50]{\includegraphics{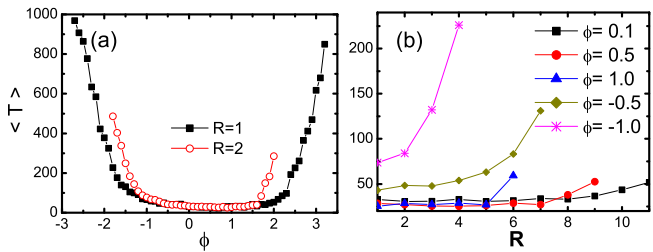}}
\caption{\label{Fig3}  (color online). Average travel time 
$\langle T \rangle$ versus $\phi$ for $R=1$ and $2$. The data 
are truncated because the system jams when $\phi$ is either 
too large or too small. The right panel shows the variation 
of $\langle T \rangle$ versus $R$ when $\phi$ is
fixed. The data are also truncated when the system jams. }
\end{figure}

Then we simulate the cars' travel time spent in the urban
transportation system. It is also an important factor for
measuring the system's efficiency. In Fig.\ref{Fig3}(a), we show the
average travel time $\langle T \rangle$ versus $\phi$ under traffic load $R=1$
and $2$. In the free-flow state, almost no congestion on vertices
occurs and the time for cars waiting in the vertex queue is
negligible, therefore, the cars' travel time is approximately equal to
their actual path length in replotted road map. But when the 
system is close to a jammed state, the travel time will increase 
rapidly. One can see that when $\phi$ is close to zero, 
the travel time is minimum.
In Fig.\ref{Fig3}(b) inset, the average travel time is much longer
when $\phi$ is negative than it is positive. These results are
consistent with the above analysis that a maximum $R_c$ occurs
when $\phi$ is slightly greater than zero. Or, in other words,
this effect can also be explained as follows: when $\phi>0$, cars
are more likely to move to the vertices with greater degree (main
roads), which enables the main roads to be efficiently used and
enhance the system's overall capability; but when $\phi$ is too
large, the main roads will more probably get jammed, and the
efficiency of the system will decrease.

Finally, we try to reproduce the fundamental diagram (flux-density
relation) of urban traffic system. It is one of the most important 
criteria that evaluates the
transit capacity for a traffic system. Our model reproduced the
phase transition and hysteresis in fundamental diagram. 

\begin{figure}
\scalebox{0.50}[0.50]{\includegraphics{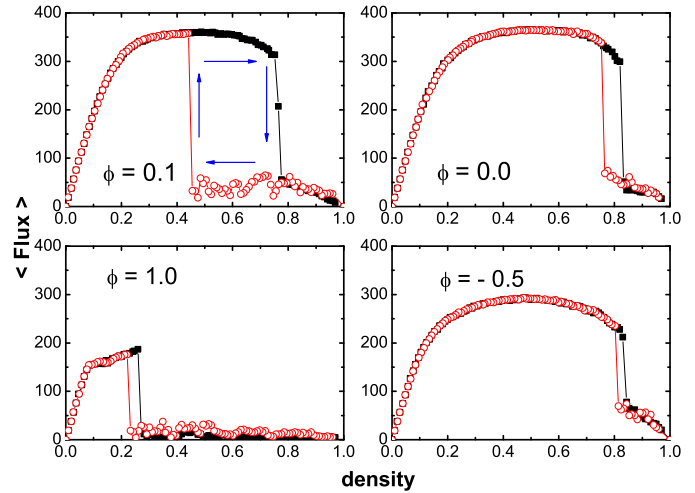}}
\caption{\label{Fig4} (color online). Fundamental diagram for a 
$N=100$ network with $m0=m=2$,$\alpha=5$, 
$\beta=0.2$, and different $\phi$. 
The data are averaged over 10 typical simulations on one 
realization of network. 
In each chart, the solid square line shows the flux variation when 
adding cars randomly to the system (increase density), while the empty 
circle line shows the flux variation when drawing out cars randomly 
from the system (decrease density). 
And the data are collected and averaged at 10,000-11,0000 steps after 
the intervention when the system has reached a steady state.
The sudden transition density values are: 0.76 and 0.45 
($\phi=0.1$), 0.82 and 0.76($\phi=0.0$), 0.26 and 0.22 ($\phi=1.0$), 
0.83 and 0.80 ($\phi=-0.5$).
For different realiazations of network, the charts are similar 
in phases, but with different transition values. }
\end{figure}

To simulate a conservative system (constant density), we count 
the number of arriving cars at each time step and add the same 
number of cars to randomly selected vertices of the system 
at the beginning of next step. 
The flux is calculated as the number of successful car turnings 
from vertex to vertex through edges
per step, as is similar to the Internet information flow. 
Here we ignore the movement of cars on a given road. 
In fact, the flux of car turnings at intersections can,
to some extent, reflect the flux on roads.
In Fig.\ref{Fig4}, the fundamental diagrams for $\phi=0.1,0.0,1.0$ 
and $-0.5$ are shown. The curves
of each diagram show four flow states: free flow, saturate flow,
bistable and jammed. For simplicity, we focus on the $\phi=0.1$
chart in the following description. As we can see, when the
density is low (less than $\approx 0.1$), all cars move freely and
the flux increases linearly with car density. It is the free-flow
state that all vertices (roads) are operated below its maximum
handling ability $C$. Then the flux's increment slows down and the
flux gradually comes to saturation ($0.10 \sim 0.45$). In this
region, the flux is restricted mainly by handling ability $C$ of
vertices. One can see that when $\phi$ is close to zero, the
saturated flux ($\approx 360$) is much 
higher than other values.

At higher density, the model reproduces an important character of
traffic flow - ``hysteresis". It can be seen that two branches of
the fundamental diagram coexist between $0.45$ and $0.76$. The
upper branch is calculated by adding cars to the system, while the
lower branch is calculated by removing cars from a jammed state
and allowing the system to relax after the intervention. In this
way a hysteresis loop can be traced (arrows in Fig.\ref{Fig4}).
The hysteresis loop indicates that the system is bistable in a
certain range of vehicle density. And as we know so far, it is the
first time that our model reproduces the hysteresis phenomenon in
scale-free network traffic and in urban network traffic.

\begin{figure}
\scalebox{0.50}[0.55]{\includegraphics{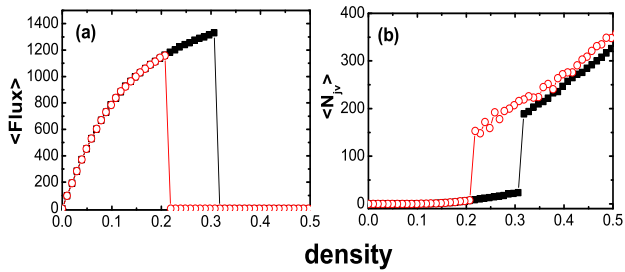}}
\caption{\label{Fig5} (color online). 
(a) Fundamental diagram for a $N=1000$ network with $m0=m=5$, 
$\alpha=1$,$\beta=0.2$ and $\phi=0.1$. (b) The averaged number 
of jammed vertices $\langle N_{jv} \rangle$. The symbols
for increasing/decreasing density are the same as in Fig.\ref{Fig4}.
One can see that the two sudden change points $0.32$ and $0.21$ 
in both charts are equal.}
\end{figure}

To test the finite-size effect of our model, we simulate some bigger 
systems with much more vertices(roads). The simulation shows 
similar phase transition and hysteresis in fundamental diagram 
as shown in Fig.\ref{Fig5}(a).

The flux's sudden drop to a jammed state from a saturated flow is a
first order phase transition. This behavior can be explained by
the sudden increment of full(jammed) vertices in the system
(See Fig.\ref{Fig5}(b)). According to the evolution rules, when a
vertex is full of cars, the cars at neighboring vertices can not
turn to it. So the cars may also accumulate on the neighboring
vertices and get jammed. This mechanism can trigger an avalanche
across the system when the car density is high. As shown in
Fig.\ref{Fig5}, the number of full vertices increase suddenly at
the same density where the flux drop to zero and almost no car can
reach its destination. As for the lower branch of the bistable 
state, starting from an initial jammed configuration, the system 
will have some jammed vertices that are
difficult to dissipate. Clearly, these vertices will decrease
the system efficiency by affecting the surrounding vertices until
all vertices are not jammed, thus we get the
lower branch of the loop.

Moreover, an important conclusion can be drawn by comparing the
$\phi=0.1$ chart with the $\phi=0.0$ chart in Fig.\ref{Fig4} 
that the $\phi=0.1$
chart has a much broader bistable region than the $\phi=0.0$ one.
This means, when the system retreats from a heavy
load jammed state, it is more difficult to reach a high
efficiency state if $\phi$ is greater than zero that cars are more
likely to move to main roads. In other words, though it
is wise to take full advantage of the main roads when the entire
traffic is light, it won't be so happy to do so at rush hours.

In conclusion, a new traffic model for scale-free networks is proposed. 
In the new perspective of mapping roads to vertices and intersections 
to edges, and incorporating road/intersection capability limits, 
the model can be applied to urban traffic system. 
In a systemic view of overall efficiency, 
the model reproduces several significant characteristics 
of network traffic, such as phase
transition, travel time, and fundamental diagram. A special
phenomenon - the ``hysteresis" - can also be reproduced.
Although the microscopic dynamics of urban traffic are not well 
captured, such as the movement of cars along streets, 
the interaction with traffic lights, and the differences 
between long and short streets, our model is still a simple and 
good representation for urban traffic, since much empirical 
evidence is well reproduced by the model. 
Further effort is deserved to consider more detailed 
elements for better mimicking real traffic systems.  
Moreover, by choosing other values of the parameters, 
the model may be applied to other networked systems, 
such as communication and power systems.

We thank Na-Fang Chu for her useful discussion and suggestion.
This work is financially supported by the National Natural Science
Foundation of China (Grant No. 10532060, 10404025) and the Australian
Research Council through a Discovery Project Grant.

\end{document}